\documentclass[showpacs,preprintnumbers,amsmath,amssymb]{revtex4}
\usepackage{graphicx,epsf}

\begin{document}
\newcommand{\be}{\begin{equation}}
\newcommand{\ee}{\end{equation}}
\newcommand{\ba}{\begin{eqnarray}}
\newcommand{\ea}{\end{eqnarray}}
\newcommand{\bk}{\mbox{\boldmath $k$}}
\newcommand{\rgl}{\rangle}
\newcommand{\lgl}{\langle}
\newcommand{\x}{\mbox{\boldmath $x$}}
\newcommand{\kay}{\mbox{\boldmath $k$}}
\newcommand{\p}{\mbox{\boldmath $p$}}
\newcommand{\arr}{\mbox{\boldmath $r$}}
\newcommand{\bu}{\mbox{\boldmath $u$}}
\newcommand{\nablab}{\mbox{\boldmath $\nabla$}}
\newcommand{\e}{\varepsilon}
\newcommand{\de}{\partial}
\newcommand{\nn}{\nonumber \\}
\newcommand{\lm}{{\ell m}}
\newcommand{\lmd}{{\ell' m'}}
\newcommand{\sqtpi}{\sqrt{\frac{2}{\pi}}}
\newcommand{\rhat}{\hat{\arr}}
\newcommand{\Ainfl}{A_{\mbox{\scriptsize inf}}}
\newcommand{\planck}{_{\mbox{\scriptsize pl}}}
\newcommand{\perms}{\mbox{perms}}
\def\simless{\mathbin{\lower 3pt\hbox
   {$\rlap{\raise 5pt\hbox{$\char'074$}}\mathchar"7218$}}}  

\def\simgreat{\mathbin{\lower 3pt\hbox  
   {$\rlap{\raise 5pt\hbox{$\char'076$}}\mathchar"7218$}}}   
\def \pom {{\hspace{ -0.1em}I\hspace{-0.2em}P}}
\def \GeV {{\rm GeV}}
\def \MeV { {\rm MeV}}
\def \mb {{\rm mb}}
\def \mub {{\rm \mu b}}

\title{
%	{\it Submitted to Phys. Rev D}\\
%	\hspace{1.in}\\
	Dynamics and Non--Gaussianity in the Weak--Dissipative Warm Inflation Scenario}

\author{S. Gupta} 
\affiliation{Institute of Cosmology and Gravitation, University of Portsmouth, Portsmouth, PO1 2EG, United Kingdom}
\email[]{sujata.gupta@port.ac.uk}

\date{\today}

\begin{abstract}

We calculate the general solutions for a warm inflationary scenario with weak dissipation, reviewing the dissipative dynamics of the two-fluid system, and calculate the bispectrum of the gravitational field fluctuations generated in the case where dissipation of the vacuum potential during inflation is the mechanism for structure formation, but is the sub--dominant effect in the dynamics of the scalar field during slow--roll. The bispectrum is non--zero because of the self--interaction of the scalar field. We compare the predictions with both those of standard, or `supercooled', inflationary models, and warm inflation models with strong dissipation and consider the detectability of these levels of non--Gaussianity in the bispectrum of the cosmic microwave background. We find that the levels of non--Gaussianity for warm and supercooled inflation are an order of magnitude different.
\end{abstract}
\maketitle
\section{Introduction}
The standard model of Hot Big Bang, or Friedmann-Robertson-Walker, cosmology with a simple, single--field {\it supercooled} inflationary phase, is highly supported by large scale structure \cite{2df}, cosmic microwave background \cite{WMAP} and weak lensing \cite{lensing} observations, where supercooled inflation describes inflation without interactions between fields. It is the most popular and developed model for use in parameter estimation, using the general assumption that the resulting perturbation spectrum is close--to scale free and Gaussian. The scenario has long been known, however, to contain a small but non--zero non--Gaussian effect due to the self--interaction of the inflaton field.  This signal was quantified in its contribution to the bispectrum statistic in \cite{ganguietal}, neglecting the effect of parametric amplification at inflation exit \cite{preheat1,preheat2}.

The challenge remains to make inflation, and cosmological parameter estimation using inflation, self--consistent, and to constrain parameters within every physically likely inflationary scenario.
 {\it Warm Inflation} \cite{ab2} describes the set of models where the presence of interactions between fields during slow--roll is included in the analysis of inflation. The dynamics and observational consequences of warm inflation have been explored in several recent papers \cite{taylber,guptaetal,brand,halletal,jpostma}. The production of radiation from the dissipation of the scalar field potential, which results from the presence of interactions, can lead to scenarios with levels of radiation density which are approximately stable. It has been shown that inflation can occur in the presence of a thermal component to the energy in the Universe. Several viable mechanisms for implementing such dissipation during inflation have been proposed in \cite{khlopov,bkep,bram,hmoss}. The popular standard, or supercooled, inflation scenario is one of the limiting cases of warm inflation. In that case there is no radiation present either due to a lack of radiation production or due to the levels of radiation produced being redshifted away in the exponential expansion of the Universe during inflation. 

A second limiting case for the warm inflation scenario is inflation with {\it strong dissipation} where the interaction terms are greater in magnitude than the Hubble friction term of standard slow--roll inflation. In \cite{ab2,bgr,bgr2,abadiab}  and \cite{guptaetal} solutions for the equation of motion of the scalar field have been calculated for the case of strong dissipation during inflation. There is a further limiting possibility out of this set of models, the {\it weak dissipation} scenario \cite{bf2,weakd,weakd2} where a constant radiation component is created from dissipation which is the sub--dominant effect in the dynamics of the inflaton during slow--roll.

In this paper we will present an analysis of the fluid dynamics of warm inflation which includes a more complete derivation of the energy constraints on warm inflation than has been previously outlined ({\it {i.e. in}} \cite{bf2}). The solutions for the equations of motion of the scalar field and its spectrum of perturbations are then calculated for warm inflation with weak dissipation for polynomial potentials. In Section {\ref{sec:sec3}} of the paper we use the solutions to provide an estimate of non--Gaussianity of the perturbations from warm inflation with weak dissipation which can be compared with the predictions for strong dissipation which are quantified in \cite{guptaetal} for various warm inflationary potentials, and with the predictions for supercooled inflation calculated in \cite{ganguietal}. Consistency tests of the form of the fluctuations generated together with accurate predictions of non--Gaussianity should be an important tool in discriminating between structure formation theories. We will conclude in Section \ref{sec:concl} with a discussion on the validity of these analyses in the wake of a recent preheating calculation for supercooled inflation \cite{preheat} which suggests that the non--Gaussianity from standard supercooled inflation has been previously greatly underestimated and may be of a level {\it already} ruled out by high resolution astrophysical data.

\section{The Background Dynamics of the 2--Fluid System During Warm Inflation}

In warm inflation, as is the case in supercooled inflation, the presence of the scalar field dominates the dynamics of the system. The energy density and energy pressure of the scalar field have the form: 
\be
         \rho_\phi=\frac{\dot{\phi}^2}{2}+V(\phi)
\label{eq:rhoback}
\ee
\be
         p_\phi=\frac{\dot{\phi}^2}{2}-V(\phi),
\label{eq:pback}
\ee
with $V\gg{\dot{\phi}}^2/2$. The potential energy of the scalar field dominates the system. The equation of motion of the scalar field can also be represented by the scalar field energy being divided into two fluids. The first fluid, representing the dynamical part of the scalar field, has an energy pressure equal to its energy density. The second fluid, representing the greater part of the energy, has an energy pressure equal to $-1 \times$ its energy density. Interactions and dissipation of the scalar field during slow--roll are, in this analysis, modelled by a constant cross--section dissipation term. So the equation of motion of the scalar field during warm inflation has the form
\be
	\ddot{\phi} + 3H\dot{\phi} + \Gamma\dot{\phi} + V'(\phi) =0,
\label{eq:infeom}
\ee
where the overdots represent time derivatives, $H=\frac{\dot{R}}{R}$
is the Hubble parameter; $R(t)$ is the cosmic expansion factor. $\Gamma$ is a {\it friction term} which incorporates the interactions of the scalar field with an environment of particles, as warm inflation models inflation occurring in a heat bath ({\it see e.g.} \cite{refpaper}). $\Gamma$ does not represent the inflaton decay rate at the end of inflation in the supercooled case, which would be modelled by rapid oscillations of the inflaton field at the base of the inflaton potential.

 To analyse the evolution of the energies in the system as a result of the constant radiation production during this scenario of inflation we require a 2-fluid model. We label the two fluids in the following calculations with the subscripts $\alpha=\{\phi,r\}$ corresponding to the scalar field energy density and the radiation energy density. The stress--energy conservation equations of the scalar field and of the radiation must contain an energy transfer term, $Q_\alpha$, between the fluids.

\be
\dot{\rho}_\alpha=-3H(\rho_\alpha +p_\alpha)+Q_\alpha ,
\label{eq:str_en_general}
\ee
such that $\sum_\alpha Q_\alpha =0$.
From Eq.({\ref{eq:rhoback}}) and Eq.({\ref{eq:pback}})
\be
\dot{\rho}_\phi=-3H\dot{\phi}^2+Q_\phi ,
\label{eq:stre_ph}
\ee
\be
= \frac{\partial\rho_\phi}{\partial \phi}\dot{\phi}.
\ee

The equation of motion Eq.(\ref{eq:infeom}) substituted into Eq.(\ref{eq:stre_ph}) together with the slow--roll assumption of inflation $\ddot{\phi}\ll\dot{\phi}$ gives the form of the energy transfer term.

Thus 
\be
\dot{\rho}_r=-4H\rho_r + \Gamma \dot{\phi}^2 .
\label{eq:str_en_general}
\ee

The evolution equation of the inflaton reduces to 
\be
	\frac{d\phi}{dt}=-\frac{1}{3H + \Gamma}\frac{dV(\phi)}{d\phi}
\label{eq:wi}
\ee
and the slow--roll condition to $(3H+\Gamma)|\dot{\phi}|\gg |\ddot{\phi}|$.

The dissipation causes radiation to be produced continuously from conversion of scalar field vacuum energy. In a preliminary work leading to the development of the warm inflation scenario \cite{bf2}, comparing temperature and thermalization scales with the Hubble scale during inflation it was shown that a dissipative component as small as $\Gamma \simgreat 10^{-5}H$  was adequate to realise $T > H$.  

We have evolved the fluid interactions of Eqs.(\ref{eq:stre_ph}-\ref{eq:str_en_general}) with a numerical simulation, where the dynamic part of the scalar field energy makes up $1\%$ of the total inflaton energy. The figure demonstrates that once the system is in the state $\dot{\rho}_r\sim 0$, it remains in this state until the slow--roll conditions are violated.

In this paper, expressions will be obtained which apply to the weak dissipative regime, $\Gamma\ll H$, so that the dominant friction term in the equation of motion is due to the Hubble expansion, and Eq.(\ref{eq:wi}) is further reduced to 
\be
\frac{d\phi}{dt}=-\frac{V'}{3H},
\label{eq:the_eq} 
\ee
while the radiation temperature resulting from weak dissipation, $T > H$, means that we are dealing with a thermalized system in every efold of inflation.

\begin{figure}
\label{onlyfig}
\begin{center}
\includegraphics[angle=0, width=70mm]{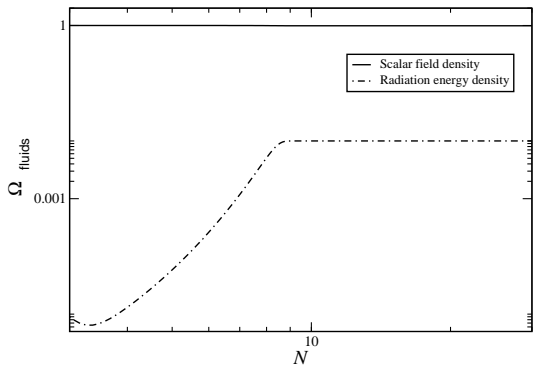}
\caption[pix]{\label{pix1} 
Evolution with efold of inflation, $N$, of scalar field $\Omega_\phi$ and radiation $\Omega_r$ energy densities for $\Gamma/H=10^{-5}$.}
\end{center}
\end{figure}

We can now evolve the behaviour of the inflaton in the weak dissipative warm inflation regime for a range of polynomial potentials. 

Focussing first on the background dynamics of the inflaton field in this regime, we start with this general form for the potential:
\be
V(\phi)=\frac{\lambda}{n!}\phi^n
\label{eq:gen_pot}
\ee

in the region $0<\phi<M$ where $M$ is $M_{\mbox{\scriptsize GUT}}=10^{14}$GeV.

Solving Eq.(\ref{eq:the_eq}) for these potentials leads to, for $n\neq 2$

\be
\phi(t)=M\left[\frac{M^{n-2}(n-2)}{(n-1)!}\frac{\lambda t}{3 H}+1\right]^{-\frac{1}{n-2}}
\label{eq:phi_neq2}
\ee
 
and for $n=2$

\be
\phi(t)=M\exp\left[-\frac{\lambda}{3H}t\right].
\label{eq:phi_eq2}
\ee

%Thus, as, during inflation $\rho_\phi\simeq V(\phi)$, for $n=2$,

%\be
%\rho(\phi)=M^2 \frac{\lambda}{2}\exp\left[\frac{-2\lambda}{3H}t \right],
%\label{eq:rho_n2}
%\ee

%and for $n\neq 2$
%\be
%\rho(\phi)=\frac{\lambda M^n}{n!}\left[\frac{(n-2)}{(n-1)!}\frac{\lambda t}{3 H}M^{n-2} +1\right]^{\frac{n}{n-2}}
%\label{eq:rho_not2}
%\ee

\section{The Evolution of Inflaton Perturbations during Warm Inflation}
\label{sec:sec3}

We assume that the production of radiation during inflation will
influence the seeds of density fluctuations, and we hold that this will apply when 
the temperature during inflation is greater than the Hubble parameter, $T > H$, and that in these bounds the thermal fluctuations of the scalar field must dominate over the quantum fluctuations.

In order to treat the fluctuations of the inflaton field
$\delta \phi({\bf x},t)$, we utilise the same stochastic approach which has been used in the calculations of non--Gaussianity which have been carried out for strong dissipative warm inflation and supercooled inflation \cite{guptaetal,ganguietal}. And also corresponding to the approach of \cite{guptaetal} and \cite{ganguietal} the effect of the self--interaction of the scalar field is treated preferentially above the back--reaction on the Hubble constant. It is assumed that the fluctuations are small --- the full inflaton field is expressed as
$\phi({\bf x},t) = \phi_0(t) + \delta \phi({\bf x},t)$
with $\phi_0$ being the homogeneous coarse--grained averaged field, $\delta \phi({\bf x},t) \ll \phi_0(t)$, the scale over which the average is taken being larger than any scale of cosmological interest.
From this the equation of motion for the full inflaton field with fluctuations emerges as
\be
	\frac{d\phi(\x,t)}{dt}=\frac{1}{3 H}
\left[e^{-2Ht}\nabla^2
\phi(\x,t) - V'(\phi(\x,t)) + \eta(\x,t) \right],
\label{eq:wi_pert}
\ee
where $\eta(\x,t)$ is a Gaussian noise term modelling {\it thermal} fluctuations with a magnitude $(k_F T/2\pi^2)^\frac{1}{2}$ \cite{abadiab} from the inflaton effectively sitting in a heat bath. $k_F$ is the wavenumber of a perturbation as it exits the inflationary causal horizon.

The fluctuation--dissipation theorem determines the properties of the noise, as in \cite{guptaetal}. With respect to physical coordinates and in momentum space,
these properties, modelled in the simplest case are
\begin{equation}
\langle \eta \rangle=0
\label{eq:noise1}
\end{equation}
\begin{equation}
\langle \eta({\bk},t) \eta({\bk}',t') \rangle=
2\Gamma T (2\pi)^3 \delta^{(3)}({\bk}- {\bk}')
\delta(t-t').
\label{eq:noise2}
\end{equation}

The choice of physical coordinates arises since we are interested
in the evolution of the inflaton mode while they are sub--horizon scale and the heat bath controls the fluctuations.

In order to calculate the predicted bispectrum from Eq.(\ref{eq:wi_pert}), the inflaton fluctuations must be expanded to first and second order:
$\delta \phi({\bf x},t)=\delta \phi_1({\bf x},t) + \delta \phi_2({\bf x},t)$, where $\delta \phi_1={\mathcal O}(\delta \phi)$ and $\delta \phi_2={\mathcal O}(\delta \phi^2)$. 

We will follow the modes through their fluctuations inside the horizon, and through freeze--out at each scale of fluctuation. On this scale comparative energetics control the scalar perturbation. We treat the physical modes adiabatically with respect to the characteristic macroscopic time scale,
the Hubble time $\sim 1/H$.  Therefore it is sensible to use physical and not comoving momenta:
\be
{\bk}_{\mbox{\scriptsize phys}} ={\bk}_{\mbox{\scriptsize com}} e^{-Ht}.
\label{eq:phys_k}
\ee
The evolution of each mode can be integrated through steps of Hubble time $\sim 1/H$, and the combination of Hubble time--steps will result in the complete solution. Hereafter for physical momenta the notation will have
no subscripts ${\bk}_{\mbox{\scriptsize phys}} \equiv {\bk}$,
and magnitudes will be denoted without boldfacing as
$k \equiv |{\bk}|$. 

Within each efold of inflation, inside the horizon:

\ba
	\frac{d}{dt}(\delta\phi_1(\kay,t))= \frac{1}{3H}[-k^2
\delta\phi_1(\kay,t) - V''(\phi_o(t))\delta\phi_1(\kay,t) \nn + \eta(\kay,t) ]
\label{eq:wisol1}
\ea
\ba
	\frac{d}{dt}(\delta\phi_2(\kay,t))= \frac{1}{3H}[-k^2
\delta\phi_2(\kay,t) - V''(\phi_o(t))\delta\phi_2(\kay,t) \nn - \frac{1}{2}V'''(\phi_o(t))\delta\phi_1(\kay,t)^2 ].
\label{eq:wisol2}
\ea

Dividing cosmic time into successive time intervals of
order $1/H$, $t_n - t_{n-1} = 1/H$, the solutions of
Eqs. ({\ref{eq:wisol1}) and (\ref{eq:wisol2}) for $t_{n-1} < t < t_n$ are respectively 
\ba
\lefteqn{\delta\phi_1(\kay,t)=} \hspace{5mm} \nn
&&A(k,t-t_{n-1})\int^t_{t_{n-1}}dt'\frac{\eta(\kay,t')}{3H}
A(k,t'-t_{n-1})^{-1} \nn &&+
A(k,t-t_{n-1})\delta\phi_1(\kay e^{-H(t_n-t_{n-1})},t_{n-1})
\label{eq:sol1}
\ea
\ba
\lefteqn{\delta\phi_2(\kay,t)=} \hspace{5mm} \nn
&&A(k,t-t_{n-1})\int^t_{t_{n-1}}dt' \nn
&& B(t')\left[\int\frac{dp^3}{(2\pi)^3}
\delta\phi_1(\p,t')\delta\phi_1(\kay-\p,t')\right]
A(k,t'-t_{n-1})^{-1} \nn 
&& + 
A(k,t-t_{n-1})\delta\phi_2(\kay e^{-H(t_n-t_{n-1})},t_{n-1}) \;,
\label{eq:sol2}
\ea
where
\ba
A(k,t)=\exp \left[-\int^{t}_{t_o}\left( \frac{k^2}{3H}+
\frac{V''(\phi_o(t'))}{3H}\right)dt' \right]
\label{eq:fn_A}
\ea
\ba
	B(t)=-\frac{V'''(\phi_o(t))}{3H}.
\label{eq:fn_B}
\ea

\section{The Warm Inflation Bispectrum}

Single field inflation models broadly predict Gaussian primordial density
fluctuations. Multiple field inflation models may lead to 
a non--Gaussian distribution (e.g. chi--squared). 
When second--order effects are taken into account, however, 
there are corrections to
these general predictions. There is a resulting non--Gaussian signal in
the cosmic microwave background radiation (CMB), the magnitude of which varies depending upon the
self-interaction of the inflaton field \cite{FRS93,luoschramm,ganguietal}.

A perfect Gaussian distribution would imply that the perturbations in the energy density have no connected correlations higher than the 2--point correlation function in real and Fourier space.

We will now calculate the {\it bispectrum} statistic, the three--point correlation function of the density perturbation distribution in Fourier space. This quantity simply translates to the harmonic bispectrum of the CMB \cite{wangkam}.

The leading order contribution to this three-point correlation
function comes from two first-order and one second-order
fluctuation as 

\ba
\lefteqn{\langle\delta\phi(\kay_1,t)\delta\phi(\kay_2,t)\delta\phi(\kay_3,t)\rangle=}
  \hspace{5mm} \nn
& &  A(k_3,t-t_{60}-1/H)\int^{t_{60}}_{t_{60-1/H}}A^{-1}(k_3,t'-t_{60}-1/H)B(t')\nn
& &  \left[\int\frac{dp^3}{(2\pi)^3}
\langle\delta\phi_1(\kay_1,t_1)\delta\phi_1(\p,t')\rangle
\langle\delta\phi_1(\kay_2,t_2)\delta\phi_1(\kay_3-\p,t')\rangle\right]\nn
& & 
+ A(k_3,t-t_{60}-1/H) 
\langle\delta\phi_1(\kay_1,t_{60})\delta\phi_1(\kay_2,t_{60})
\nn
& & \delta\phi_2(\kay_3e^{-1},t_{60}-1/H)\rangle
\nn 
&&  +(\kay_1 \leftrightarrow \kay_3))
+(\kay_2 \leftrightarrow \kay_3)).
\label{eq:inf_bisp}
\ea

Thus Eq. (\ref{eq:inf_bisp}) becomes
\ba
\lefteqn{\langle\delta\phi(\kay_1,t)\delta\phi(\kay_2,t)\delta\phi(\kay_3,t)\rangle
\approx B(t_{60}) \Delta t_F} \hspace{5mm} \nn
&& \left[\int\frac{dp^3}{(2\pi)^3}\langle\delta\phi_1(\kay_1,t_1)\delta\phi_1(\p,t')\rangle\langle\delta\phi_1(\kay_2,t_2)\delta\phi_1(\kay_3-\p,t')\rangle \right. \nn
&& \left. +(\kay_1 \leftrightarrow \kay_3)) + (\kay_2 \leftrightarrow \kay_3)
\right].
\label{eq:infbisp2}
\ea

It is possible to write a general expression for the bispectrum for
slow roll, single field, supercooled inflation models as well as for
the set of warm inflation models in terms of products of power spectra and a prefactor, $\Ainfl$, which depends upon the interactions during inflation, on the shape of the inflationary potential and upon the shape of the triangles used to measure the bispectrum statistic.
\ba
\lefteqn{\langle\Phi(\kay_1)\Phi(\kay_2)\Phi(\kay_3)\rangle=}\hspace{5mm}\nn
&& \Ainfl (2\pi)^3\delta^3(\kay_1+\kay_2+\kay_3) \left[P_\Phi(\kay_1)P_\Phi(\kay_2)+\perms\,\right]
\label{eq:bisp_a_infl}
\ea
The relation between the scalar field fluctuation and the
gravitational field has the simple form (from \cite{bard})

\be
\Phi(\kay)=-\frac{3}{5}\frac{H}{\dot{\phi}}\delta \phi(\kay),
\label{eq:bardeen}
\ee
thus $\Ainfl$ calculated using equilateral bispectrum triangles for a weakly dissipative warm inflation regime is
\ba
\label{eq:ainfl_wi}
\Ainfl^{\mbox{\scriptsize weak}}=-\frac{10}{3}\left(\frac{\dot{\phi}}{H}\right)
\left[\frac{V'''(\phi_o(t_F))}{3H}\right] \;,
\ea
which, plugging in the temperature $T$ calculated using the relation
$\rho_r=\left(g_\ast\pi^2/30\right)T^4$, where $g_\ast$ is the number of relativistic fields $\sim 150$, and taking into account the 
relation between the radiation energy density and the scalar field
potential $\rho_\phi\simeq V(\phi)$, given in Eq.(\ref{eq:gen_pot}), and the CMB amplitude, given by 
\be
\delta_H = \frac{2}{5}\frac{H}{\dot{\phi}}\delta \phi \;,
\label{eq:del_h} 
\ee
which, from CMB results of the Wilkinson Microwave Anisotropy Probe, is
$\delta_H = 1.94 \times 10^{-5}$ \cite{smoot_isot,bunlidwhit,bunnwhite}, gives for a quartic potential a value of $\Ainfl ^{\mbox{\scriptsize weak}}=7.65\times 10^{-3}$. Our derivation of  $\Ainfl$ mirrors the derivation of this value, the coefficient of the bispectrum of the scalar field gravitational potential, and also the coefficient of the bispectrum of the CMB in the Sachs--Wolfe region, of \cite{guptaetal}.

\section{Conclusions}
\label{sec:concl}

We have followed the behaviour of the inflaton field and of radiation during warm inflation in the limit of weak dissipation. We have evolved the scalar field perturbation up to second order and calculated the non--Gaussian signal in the bispectrum statistic which results from the self--interaction of the scalar field. The value we obtain  $\Ainfl ^{\mbox{\scriptsize weak}}=7.65\times 10^{-3}$ is an order of magnitude smaller than the corresponding signal for a quartic inflationary potential for the strong dissipative and cool inflationary scenarios \cite{guptaetal,ganguietal}.

At the time of the publication of \cite{guptaetal} it was believed by the authors that the difference in the prediction for the bispectrum statistics of warm inflation with strong dissipation: $\Ainfl ^{\mbox{\scriptsize strong}}=7.44\times 10^{-2}$; and the prediction for supercooled inflation:  $\Ainfl ^{\mbox{\scriptsize cool}}=5.56\times 10^{-2}$ \cite{ganguietal}; would be overwhelmed by a contribution from general relativistic second order perturbation theory of $\Ainfl\sim 1$ \cite{pynecarr}. However, an energetic analysis of the evolution of second order perturbation theory \cite{malikwands} has since shown that this is not the case.

However, one incentive for the use of the warm inflation model over the supercooled inflation model, the implications of the preheating mechanism on the radiation perturbations, has a new possible observational implication. In \cite{preheat} it is demonstrated that parametric resonance during preheating can enhance the  non-Gaussianity such that $\Ainfl^{\mbox{\scriptsize cool}}$ can easily become as large as $\sim 100$. The implication of this for the result of this paper is that the difference between the predictions of Gaussianity for strongly and for weakly dissipative warm inflation may be both significant and observable in the future. Although this will not be distinguishable with the simple observation of the bispectrum with satellite CMB data, the best predicted constraint on this from the upcoming Planck satellite being $\Delta\Ainfl=10$ \cite{komsperg}, the difference between the non--Gaussian signals from the strongly and weakly dissipative warm inflation scenarios might be tested for with others of the many non--Gaussianity statistics i.e. the trispectrum \cite{trisp} and the two--point correlation of peaks \cite{peaks}. 

\section{Acknowledgements}
Sujata Gupta is funded by PPARC.

\vspace{0.5cm}

\end{document}